\documentclass[prb,superscriptaddress,twocolumn,floatfix]{revtex4-1}

\usepackage[FIGTOPCAP]{subfigure}
\usepackage[usenames]{color}
\usepackage{amsmath,amssymb}
\usepackage[pdftex]{hyperref,graphicx}
\hypersetup{colorlinks = true, urlcolor = blue, linkcolor = blue, citecolor = blue}

\begin{document}
\title{Spin-valley locked instabilities in moir\'e transition metal dichalcogenides with conventional and higher-order Van Hove singularities}
\author{Yi-Ting~Hsu}
\email{yhsu2@nd.edu}
\affiliation{Department of Physics, University of Notre Dame, Notre Dame, IN 46556, USA}
\affiliation{Condensed Matter Theory Center and Joint Quantum Institute, University of Maryland, College Park, MD 20742, USA}
\author{Fengcheng Wu}
\affiliation{Condensed Matter Theory Center and Joint Quantum Institute, University of Maryland, College Park, MD 20742, USA}
\affiliation{School of Physics and Technology, Wuhan University, Wuhan 430072, China}
\author{S. Das Sarma}
\affiliation{Condensed Matter Theory Center and Joint Quantum Institute, University of Maryland, College Park, MD 20742, USA}
\date{\today}

\begin{abstract}
Recent experiments have observed correlated insulating and possible superconducting phases in twisted homobilayer transition metal dichalcogenides (TMDs). Besides the spin-valley locked moir\'e  bands due to the intrinsic Ising spin-orbit coupling, homobilayer moir\'e  TMDs also possess either logarithmic or power-law divergent Van Hove singularities (VHS) near the Fermi surface, controllable by an external displacement field. The former and the latter are dubbed conventional and higher-order VHS, respectively. Here, we perform a perturbative renormalization group (RG) analysis to unbiasedly study the dominant instabilities in homobilayer TMDs for both the conventional and higher-order VHS cases. We find that the spin-valley locking largely alters the RG flows and leads to instabilities unexpected in the corresponding extensively-studied graphene-based moir\'e  systems, such as spin- and valley-polarized ferromagnetism and topological superconductivity with mixed parity. In particular, for the case with two higher-order VHS, we find a spin-valley-locking-driven metallic state with no symmetry breaking in the TMDs despite the diverging bare susceptibility. Our results show how the spin-valley locking significantly affects the RG analysis and demonstrate that moir\'e  TMDs are suitable platforms to realize various interaction-induced spin-valley locked phases, highlighting physics fundamentally different from the well-studied graphene-based moire systems.
\end{abstract}

\maketitle
\section{Introduction}
Twisted bilayer Van der Waals materials have been receiving extensive attention following the discoveries of superconducting and correlated phases with a rich variety of spontaneously broken symmetries and topological properties\cite{Exp_SC_BLG_Cao2018,Exp_CI_BLG_Cao18,Exp_BLG_sc_Dean,Exp_BLG_nematic,Exp_BLG_nematic_nphys,
Exp_FM_BLG,Exp_BLG_LinearT_Young,Exp_BLG_Chern_Barcelona,Exp_BLG_compressibility_PRL,BLG_QAHE_Serlin2020,
Exp_WSe2bilayer_Dean2020,Exp_WSe2bilayer_STM_NPhys,Exp_WSe2WS2_CIWigner_Wang2020,Tang2020,
Exp_WSe2WS2_cascadeCI_Shan2020,Exp_WSe2WS2_Stripe_Mak2020}.  
In particular, group-VI monolayer transition metal dichalcogenides (TMD) have become attractive building blocks for moir\'e systems for their large spin-orbit coupling allowed by the broken inversion symmetry\cite{TMD_Xiao_PRL} and their experimental tunabilitiy. 
This valley-dependent spin-orbit coupling acts like an effective Zeeman field with opposite out-of-plane directions in the two valleys, which leads to an effective locking between the spin and valley degrees of freedom\cite{Suzuki2014,TMD_Xiao_PRL}. 
Such spin-valley locking, which carries over to twisted bilayer TMDs, not only allows optical controls of valley degrees of freedom\cite{Cao2012,Mak2012,Jones2013} but could also lead to exotic topological and symmetry broken phases\cite{Hsu2017,BilayerTMD_Fu}.   
In fact, several recent experiments on twisted hetero- or homobilayer TMDs have reported correlated insulating states at multiple fractional fillings\cite{Exp_WSe2WS2_CIWigner_Wang2020,Exp_WSe2WS2_cascadeCI_Shan2020,
Exp_WSe2WS2_Stripe_Mak2020,Exp_WSe2WS2_scanningimpedence_CIs_Cui2020} and metallic phases at small dopings away from these insulating phases\cite{moire_metal_Pan,Exp_QC_moireTMD,Exp_MottTransMoire_Mak}, as well as possible nearby superconductivity in homobilayer WSe$_2$\cite{Exp_WSe2bilayer_Dean2020}. This spin-valley locking feature qualitatively distinguishes the TMD moire systems from the well-studied graphene moire systems where one must deal with spin, valley, and sublattice symmetries, all of which in the single-particle graphene Hamiltonian. 

\begin{figure}[t]
\includegraphics[width=8cm]{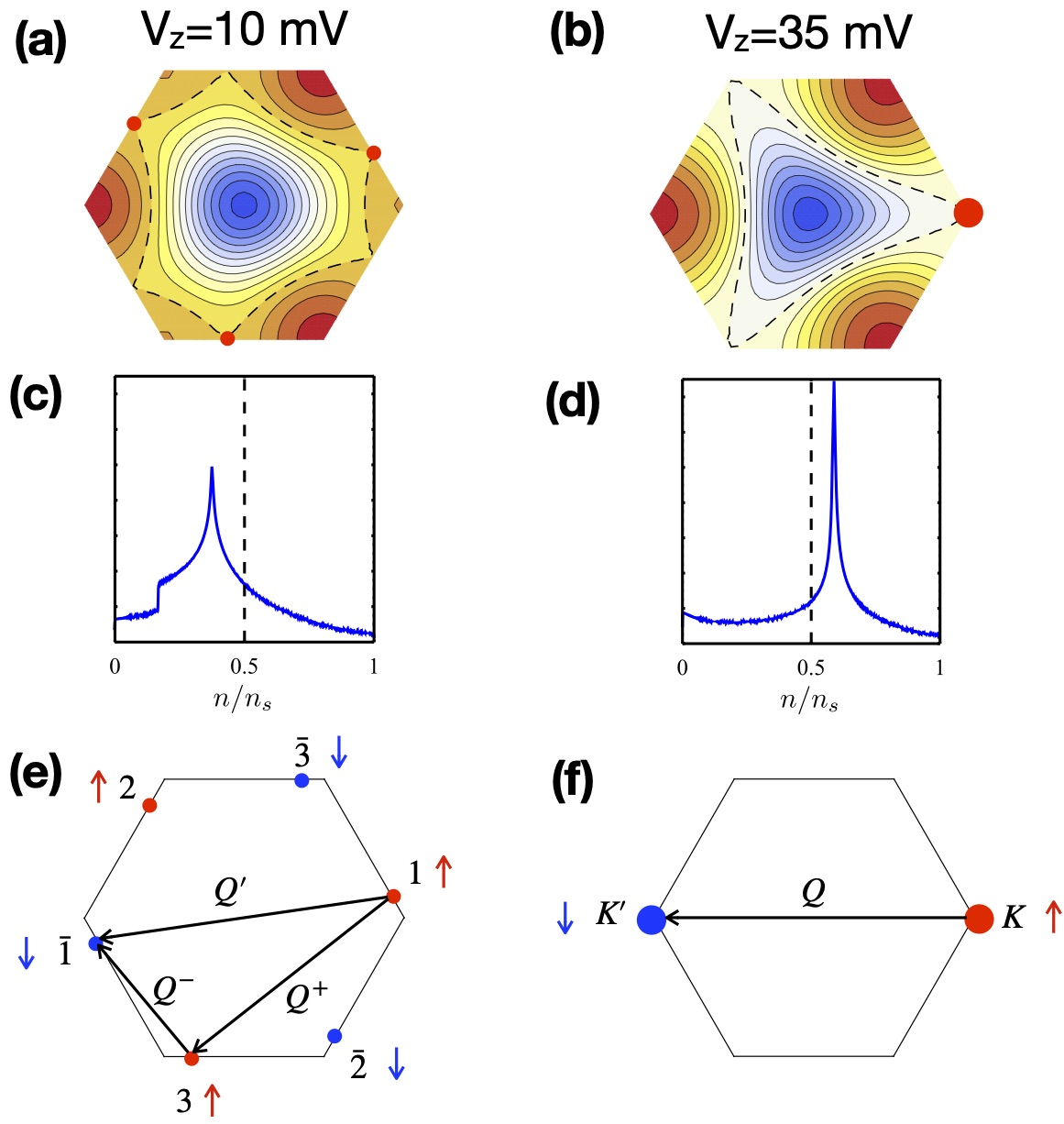}
\caption{(a) Energy contour plot for the first moir\'e conduction band in $K$ valley of twisted bilayer TMD under a layer dependent potential $U=10$ meV that is generated by an out-of-plane displacement field. (b) Similar as (a) but for $U=35$ meV. The dashed lines and the red dots in (a) and (b) mark the Fermi surfaces at the van Hove filling and the van Hove points, respectively. (c) The calculated density of states as a function of filling factor for the bands shown in (a). The peak results from the three conventional van Hove singularities shown in (a). (d) The calculated density of states as a function of filling factor for the bands shown in (b). The peak results from the higher-order van Hove singularity shown in (b). (e) Schematics of the six-patch model we consider for the representative van Hove fermiology in (a) with conventional VHS. (f) Schematics of the two-patch model we consider for the van Hove fermiology in (b) with higher-order VHS. In (e) and (f), the hexagon, red points, and blue points represent the mBZ, the patch centers from $K$ valley, and those from $K'$ valley, respectively. The arrows represent the characteristic momenta connecting the inter- and intra-valley patches. (a)-(d) are taken from Ref. \onlinecite{Thy_bilayerWSe2_dispersion} with permission.}
\label{FS}
\end{figure}

Besides the spin-valley locking, another feature of homobilayer TMD is that there are two types of van Hove singularities (VHS) near the Fermi level that are tunable by an external displacement field\cite{Thy_bilayerWSe2_dispersion}. 
The first type is the conventional VHS, where the density of states is logarithmically divergent such that only the Cooper instability diverges as log square unless the Fermi surface is perfectly nested\cite{Schulz_RG_1987}. 
The second type is the higher-order VHS\cite{HOVH_dispersion}, where the density of states has a stronger power-law divergence such that the bare susceptibilities in both the Cooper and particle-hole channels diverge with power-law as well\cite{supermetal,RG_HOVH_Scherer,RG_HOVH_Nandkishore} [see Appendix \ref{app:bubbles2}]. 
Homobilayer TMD under a general displacement field has six conventional VHS, three from each valley (spin), located at field-tunable positions on the moir\'e  Brillouin zone (mBZ) boundary [see Fig. \ref{FS}(a)(c)]. 
Nonetheless, at certain field strength, the three conventional VHS from the same valley merge into one higher-order VHS with an exponent $\eta=1/3$ at the valley center [Fig. \ref{FS}(b)(d)]. 
We note that whether the VHS is conventional with logarithmic divergence or higher-order with power law divergence is physically controllable by an applied external displacement field, and is therefore of experimental relevance.  The possibility of the higher-order power-law VHS is another important feature distinguishing TMD from graphene systems. 

Here, we investigate the dominant Fermi surface instabilities in homobilayer TMD for cases with six conventional VHS or two higher-order VHS at the Fermi level. 
The method we adopt is a perturbative renormalization group (RG) approach dubbed parquet RG\cite{Schulz_RG_1987,Salmhofer_RG_1998,RG_FeSc_Chubukov_PRB,RG_FeSc_Fernandes_PRX,RG_oddparitysc_Yao,Nphys_Nandkishore2012}, which can reliably predict phase diagrams when the correlated gaps are smaller than the band width (i.e. weak coupling), and has been applied to various graphene-based moir\'e  systems\cite{RG_Moire_Fu_PRX,PatchRG_nonesting_PRB,RG_Moire_Nandkishore,DBLG_RG_Hsu,RG_HOVH_Scherer,RG_HOVH_Nandkishore,NematicSc_TBG_Chubukov,HexSC_Nandkishore}.   
Given that the density of states are mainly contributed by the VHS, we consider only patches centered at the VH points instead of the full moire Brillouin zone (mBZ). The parquet RG formalism then allows us to unbiasedly treat both the particle-hole and particle-particle instabilities driven by the intra- and inter-patch interactions. 
This RG method has been shown to approximate the direct diagrammtic results qualitatively for both cases with conventional\cite{PhysRevB.64.165107} and higher-order VHS\cite{RG_HOVH_Scherer}, and can provide precise experimental information about the dominant interaction-induced instabilities and the associated quantum phase transitions.

We find that the spin-valley locking significantly alters the RG flows for both the conventional VHS and higher-order VHS cases, and leads to phases unexpected in graphene-based moir\'e  systems.  
In particular, the important differences between the RG analyses for spin-degenerate and spin-valley locked bands are that (1) the inter- and intra-patch interactions are subjected to extra constraints from fermionic statistics in the latter case, and (2) the fermion flavor reduces from two to one. Due to these differences, in the six-patch conventional VHS case, we find that the spin- and valley-polarized phase and mixed-parity topological superconductivity are energetically favorable depending on whether the bare inter- and intra-patch interactions are repulsive or attractive. In contrast, the pair-density wave and all the charge and magnetic instabilities with well-defined spin or valley characters, which were found in moir\'e  systems with six spin-degenerate VH patches by the same method\cite{DBLG_RG_Hsu}, are now suppressed by the spin-valley locking. 

As for the two-patch higher-order VHS case, the particle-hole nesting degree within the two patches plays a crucial role under the symmetry contraints imposed by the spin-valley locking.  
Specifically, in the perfect nesting limit, we find a metallic state without symmetry breaking up to second-order perturbations due to the marginal inter-valley density-density interaction. 
When the nesting degree deviates from perfect, we find another metallic state without symmetry breaking when the bare density-density interaction is repulsive. 
This metallic state is similar to the ``supermetal'' state found in systems with one\cite{supermetal} or three\cite{RG_HOVH_Scherer} spin-degenerate higher-order VHS in that no long-range order is developed at low temperatures despite that the bare susceptibilities are diverging.
Nonetheless, the two metallic states are different in that instead of flowing to a finite fixed point, here the interaction that drives the particle-hole instabilities becomes irrelevant in the low-energy limit. 
The lack of symmetry breaking in our case is a consequence the spin-valley locking, which suppresses the influence from other interactions that are allowed in the spin-degenerate case. 
Another difference from the spin-degenerate case with a single higher-order VHS\cite{supermetal} is that when the bare density-density interaction is attractive, we find that a mixed-parity superconducting state is dominant. 
We note that recently metallic states have been reported in TMD moire systems by two groups\cite{Exp_QC_moireTMD,Exp_MottTransMoire_Mak}. 

The rest of the paper is structured as follows. In section II, we discuss the non-interacting dispersions of twisted homobilayer TMDs, emphasizing specifically the two types of van Hove singularity patterns together with their density of states, and the corresponding six- and two-patch models. In section III-V, we show the renormalization group (RG) calculation step by step respectively for the six-patch and two-patch models, including the non-interacting susceptibilities, RG equations for the inter- and intra-valley interactions, and the tendencies for the symmetry-allowed instabilities. In section VI, we present the RG flows and the resulting phase diagrams for the six- and two-patch models. Finally, we discuss the experimental relevance of our results in section VII.

\section{The two types of Van Hove patterns}\label{sec:model}
We start by considering the model $H_0=\sum_{s=\uparrow,\downarrow}H_{0}^s$ for the topologically trivial first moir\'e  valence band in a homobilayer TMD\cite{Thy_bilayerWSe2_dispersion,PhysRevLett.121.026402,PhysRevLett.122.086402}. Due to the spin $s_z$-preserving spin-orbit coupling, the spin labels $s=\uparrow$ and $\downarrow$ are tied to the valley labels $\tau=K$ and $K'$, respectively, and the two Hamiltonians $H_{0}^\uparrow$ and $H_{0}^\downarrow$ are related by the time-reversal symmetry.
 
In the presence of a small out-of-plane displacement field, there are three inequivalent conventional VHS per spin (valley) on the mBZ boundary below half hole filling [see Fig. \ref{FS}(a)]. As we increase the displacement field, the locations of the three VHS with spin $\uparrow$ ($\downarrow$) will move along the mBZ boundary towards the mBZ corner $\kappa_{+}$ ($\kappa_{-}$) and merge into a single higher-order VHS at some critical field strength [see Fig. \ref{FS}(b)]. 
The key difference between the two cases, besides the number of VHS, is that the density of states in the former six-VHS case are logarithmically divergent [see Fig. \ref{FS}(c)] and that in the latter two-VHS case are power-law divergent with an exponent $\eta=1/3$ [see Fig. \ref{FS}(d)]. 

In the six-VHS case, the low-energy dispersions near these van Hove points $\textbf{P}_n$, $n=1,2,3$ from the spin-up Hamiltonian $H_0^{\uparrow}$ are given by 
\begin{align}
&\epsilon_{\textbf{k}}^1 \approx \sum_{\alpha=x,y} \sum_{\beta=x, y} w_{\alpha \beta} (\textbf{k}-\textbf{P}_1)_\alpha(\textbf{k}-\textbf{P}_1)_\beta\nonumber\\
&\epsilon_{\textbf{k}}^2 =  \epsilon_{\hat{\mathcal{R}}_3^{-1}\textbf{k}}^1, \,\,\, \epsilon_{\textbf{k}}^3 =  \epsilon_{\hat{\mathcal{R}}_3 \textbf{k}}^1, 
\label{eq:dispersion1}
\end{align} 
where both $\textbf{k}$ and the VH point positions $\textbf{P}_1$, $\textbf{P}_2=\hat{\mathcal{R}}_3 \textbf{P}_1$, $\textbf{P}_3 = \hat{\mathcal{R}}_3^{-1} \textbf{P}_1$ are measured relative to the mBZ center $\bar{\Gamma}$ point with $\hat{\mathcal{R}}_3$ being the $+2\pi/3$ rotation matrix.
Here, we keep only up to the quadratic terms in momentum $\textbf{k}$. 
Both the VH point positions $\textbf{P}_n$ and the coefficient matrix $w$ are tunable by the displacement field, where $w$ is a symmetric real matrix. Since $w$ describes the dispersion around a saddle point, $w$ obeys $\text{Det} (w) <0$. The dispersions of the spin-down Hamiltonian $H_0^{\downarrow}$ are given by Eq. \ref{eq:dispersion1} as well but with van Hove points $\textbf{P}_{\bar{n}}=-\textbf{P}_n$, $n=1,2,3$.   

When these six VHS are at the Fermi level, since the density of states are mainly from the portion of FS near these VHS, we can make the ``patch approximation'' and consider only momentum-space patches centered at the VH points \textbf{P}$_n$ and \textbf{P}$_{\bar{n}}$ with a patch size $k_\Lambda$. We focus on the weak-coupling regime in which the ultraviolet energy cutoff $\Lambda$ that corresponds to the patch size is small compared to the band width.  
This resulting six-patch single-particle model is very similar to that for the spin-degenerate twisted double bilayer graphene considered in Ref. \onlinecite{DBLG_RG_Hsu} except the spin-valley locking. 

For the two-VHS case, the low-energy dispersion near the VH points from the two valleys are given by 
\begin{align}  
&\epsilon_{\textbf{k}}^K=\kappa(k_x^3-3k_xk_y^2)\nonumber\\
&\epsilon_{\textbf{k}}^{K'}=-\kappa(k_x^3-3k_xk_y^2),
\label{eq:dispersion2}
\end{align}
where the momenta $\textbf{k}$ are measured relative to the mBZ corners, and $\kappa$ is given by the overall energy scale. Importantly, the Fermi surface near the two higher-order van Hove points is perfectly nested at momentum $\textbf{q}=2\textbf{Q}$ since the dispersions satisfy $\epsilon_{\textbf{k}}^K=-\epsilon_{\textbf{k}}^{K'}$ [see Fig. \ref{FS}(b)]. Moreover, the density of states near these higher-order VH points exhibit power-law divergence as [see Appendix \ref{app:bubbles2} and Fig. \ref{FS}(d)]  
\begin{align}
\nu(E)=\bar{\nu}|E|^{-\eta},
\end{align}
where $\eta=1/3$ and $\bar{\nu}=\frac{1}{4\sqrt{3}\pi^{3/2}}\frac{\Gamma(1/3)}{\Gamma(5/6)}|\kappa|^{-2/3}$.  

\section{Bare susceptibilities}
We now perform RG analyses for both the six-patch model with six conventional VHS and the two-patch model with higher-order VHS. The first step is to study the inter- and intra-patch non-interacting static susceptibilities in the particle-hole and particle-particle channels 
\begin{align}
&\Pi_{\rm ph}^{nm}(\textbf{q})=-\int d\textbf{k}\frac{f_{\epsilon^n_{\textbf{k}}}
-f_{\epsilon^m_{\textbf{k}+\textbf{q}}}}{\epsilon^n_{\textbf{k}}-\epsilon^m_{\textbf{k}+\textbf{q}}}\nonumber\\
&\Pi_{\rm pp}^{nm}(\textbf{q})=\int d\textbf{k}\frac{1-f_{\epsilon^n_{\textbf{k}}}
-f_{\epsilon^m_{-\textbf{k}+\textbf{q}}}}{\epsilon^n_{\textbf{k}}+\epsilon^m_{-\textbf{k}+\textbf{q}}}, 
\label{eq:bubbles} 
\end{align}
where $n$ and $m$ are patch labels. For the six-patch case, there are four inequivalent susceptibilities per channel at momenta $\textbf{q}=0,\textbf{Q}'_n,\textbf{Q}^+_{nm},$ and $\textbf{Q}^-_{n\bar{m}}$, respectively. The latter three momenta connect patch $n$ with the opposite patch $\bar{n}$, another patch in the same valley $m\neq n$, and another patch in the opposite valley $\bar{m}\neq\bar{n}$, respectively [see Fig. \ref{FS}(e)]. 

Among these eight bare susceptibilities, the density of states $\Pi_{\rm ph}^{nn}(0)$ and the Cooper instability $\Pi_{\rm pp}^{n\bar{n}}(0)$ exhibit logarithmic and logarithmic square divergence,  respecitvely, as
\begin{align}
&\Pi_{\rm ph}^{nn}(0)=\nu_0\ln\frac{\Lambda}{\text{max}(T,|\mu|)}\nonumber\\
&\Pi_{\rm pp}^{n\bar{n}}(0)=\frac{\nu_0}{2}\ln\frac{\Lambda}{\text{max}(T,|\mu|)}\ln\frac{\Lambda}{T}, 
\label{DOScooper6}
\end{align}
where $\nu_0$ depends on the specific dispersions in Eq. \ref{eq:dispersion1}, $\mu$ is the chemical potential, $T$ is the temperature, and $\Lambda$ is the ultraviolet cutoff of the patch model. 

Given that the Fermi surface (FS) is weakly nested in general between both the same-spin and the opposite-spin patches for twisted bilayer TMDs [see Fig. \ref{FS}(a)], we parametrize the corresponding susceptibilities using the density of states as 
\begin{align} 
&\Pi_{\rm ph}^{n\bar{m}}(\textbf{Q}^-_{n\bar{m}})=\gamma_{ph}^-\Pi_{\rm ph}^{nn}(0),
~~~\Pi_{\rm ph}^{n\bar{n}}(\textbf{Q}'_n)=\gamma_{ph}'\Pi_{\rm ph}^{nn}(0),\nonumber\\
&\Pi_{\rm ph}^{nm}(\textbf{Q}^+_{nm})=\gamma_{ph}^+\Pi_{\rm ph}^{nn}(0), 
~~~\Pi_{\rm pp}^{nm}(\textbf{Q}^-_{\bar{n}m})=\gamma_{pp}^-\Pi_{\rm ph}^{nn}(0),\nonumber\\    
&\Pi_{\rm pp}^{nn}(-\textbf{Q}'_n)=\gamma_{pp}'\Pi_{\rm ph}^{nn}(0),
~~~\Pi_{\rm pp}^{n\bar{m}}(\textbf{Q}^+_{\bar{n}\bar{m}})=\gamma_{pp}^{+}\Pi_{\rm ph}^{nn}(0),  
\label{nesting}
\end{align}
where patch $n$ and $m\neq n$ belong to the same valley, and patch $\bar{n}$ is the opposite patch belonging to the other valley. Here, $\gamma_{ph}^-,\gamma_{ph}'\geq0$ ($\gamma_{pp}^-,\gamma_{pp}'\geq0$) characterize the nesting degrees in the particle-hole (particle-particle) channel between different inter-valley pockets, and $\gamma_{ph}^+\geq0$ ($\gamma_{pp}^+\geq0$) characterizes those between intra-valley pockets. These nesting parameters are in principle not bounded by unity. 

For the two-patch case, by using the dispersions $\epsilon^K_{\textbf{k}}$ and $\epsilon^{K'}_{\textbf{k}}$ in Eq. \ref{eq:dispersion2} to compute the susceptibilities in Eq. \ref{eq:bubbles}, we find that both the density of states and the Cooper instability are power-law divergent [see Appendix \ref{app:bubbles2}]  
\begin{align}
&\Pi_{\rm ph}(0)= \bar{\nu}_{ph}T^{-1/3}\nonumber\\
&\Pi_{\rm pp}(0)= \bar{\nu}_{pp}T^{-1/3},  
\label{DOScooper2}
\end{align}
where $\bar{\nu}_{ph}=\frac{\bar{\nu}}{4}\int d\epsilon|\epsilon|^{-1/3}\cosh^{-2}(\epsilon/2)\sim 1.14\bar{\nu}$ and $\bar{\nu}_{pp}=\frac{\bar{\nu}}{2}\int d\epsilon|\epsilon|^{-4/3}\tanh(\epsilon/2)\sim 3.4\bar{\nu}$. 
Moreover, due to the perfect nesting between the Fermi surface within the two patches [see Eq. \ref{eq:dispersion2} and Appendix \ref{app:bubbles2}], the particle-hole susceptibility at momentum $\textbf{q}=\textbf{Q}$ [defined in Fig. \ref{FS}(f)] also exhibits power-law divergence and with the same coefficient as the Cooper instability 
\begin{align}
&\Pi_{\rm ph}(\textbf{Q})= \Pi_{\rm pp}(0).   
\label{nesting2}
\end{align}
These bare susceptibilities turn out to be the relevant ones that determine the dominant instabilities in the two-patch case, as we will show later. 

\begin{figure}[t]
\includegraphics[width=8cm]{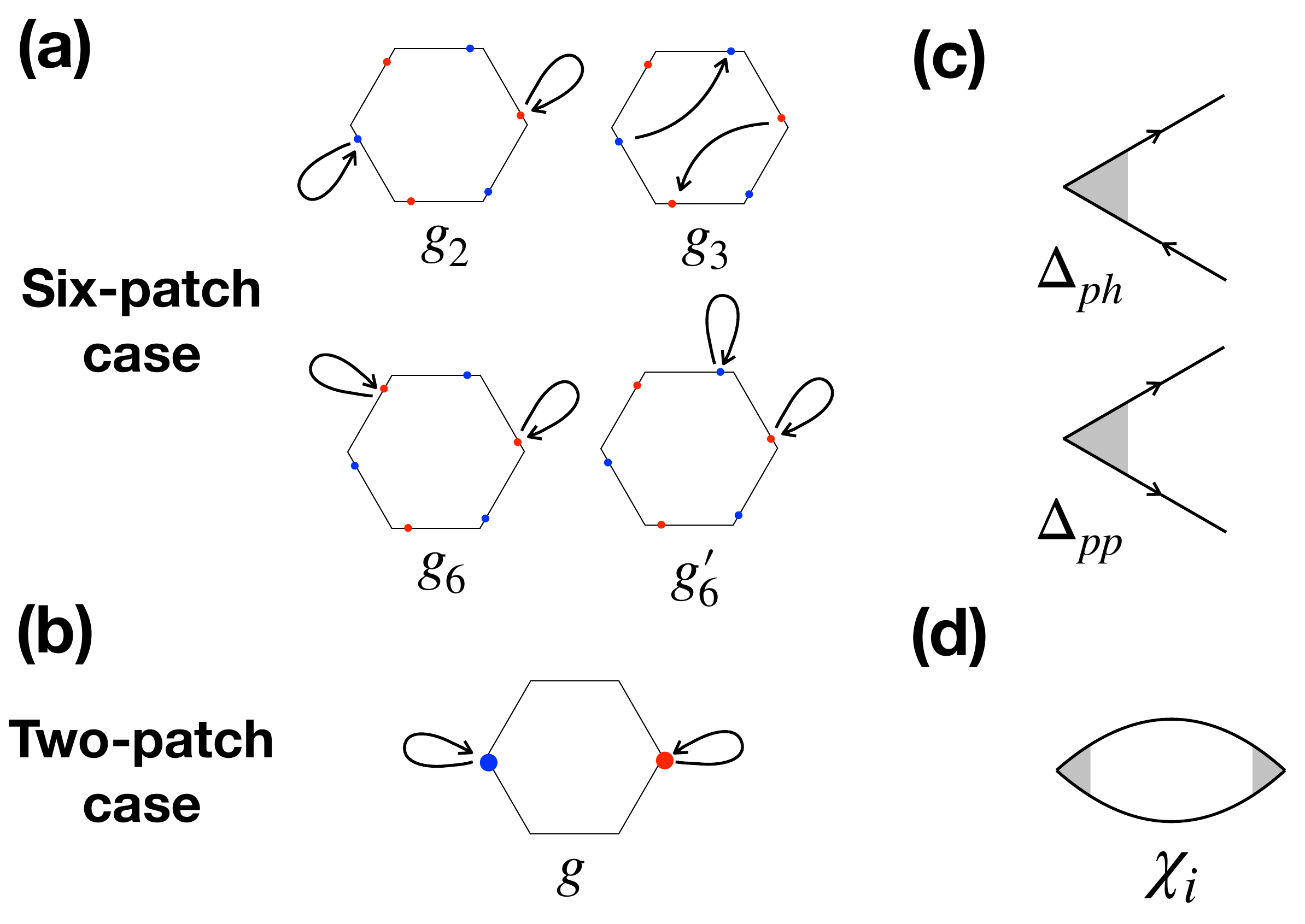}
\caption{(a)Schematics for all the momentum-preserving interactions allowed under the spin-valley locking for (a) the six-patch model and (b) the two-patch model. The hexagon, red dots, and blue dots represent the mBZ, the VH points from valley $K$, and those from valley $K'$, repectively. Diagrammatic expressions for (c) the test vertices in the particle-hole and particle-particle channels and (d) the susceptibilities.}
\label{gidf}
\end{figure}
\section{Inter-patch effective interactions}
Equipped with the bare susceptibilities, we now discuss the inter- and intra-patch interactions for both the six-patch and two-patch models. 
\subsection{Six-patch model}
For the six-patch case, there are only four inequivalent interactions allowed by momentum and spin-$s_z$ conservation under the spin-valley locking [see Fig. \ref{gidf}(a) for schematics] 
\begin{align}
H^{(6)}_{\text{int}}&=
\sum_{n=1}^3
~\tilde{g}_2\psi^{\dagger}_{n}\psi_{n}\psi^{\dagger}_{\bar{n}}\psi_{\bar{n}}+\sum_{n=1}^3\sum_{m\neq n}[~\tilde{g}_3\psi^{\dagger}_{m}\psi^{\dagger}_{\bar{m}}\psi_{\bar{n}}\psi_{n}\nonumber\\
&+\frac{1}{2}\tilde{g}_6(\psi^{\dagger}_{n}\psi_{n}\psi^{\dagger}_{m}\psi_{m}
+\psi^{\dagger}_{\bar{n}}\psi_{\bar{n}}\psi^{\dagger}_{\bar{m}}\psi_{\bar{m}})
+\tilde{g}_6'\psi^{\dagger}_{n}\psi_{n}\psi^{\dagger}_{\bar{m}}\psi_{\bar{m}}~], 
\label{eq:H6}
\end{align}
where $\psi_{n}$ annihilates electrons on patch $n=1,2,3$ from valley $K$, $\bar{n}$ labels the patch from valley $K'$ that centers at the opposite momentum to patch $n$, and $m\neq n$ labels the rest of the valley-$K$ patches besides patch $n$. Since the spin and valley degrees of freedom are locked, electrons from valley $K$ and $K'$ have up- and down-spin, respectively. Here, we assume the interactions are independent of the momentum difference within a patch, and only depend on the momentum difference among patches.

Among these four interactions, the $\tilde{g}_6$ term is the density-density interaction between same-valley patches, the $\tilde{g}_2$ and $\tilde{g}_6'$ terms are those between opposite-valley patches, and the $\tilde{g}_3$ term is a zero-momentum (BCS) pair scattering process with an intra-valley momentum transfer $\textbf{Q}^+$. 
Importantly, due to the spin-valley locking, the inter-valley exchange scatterings are forbidden since they cause spin flips. Moreover, since there is only one spin species per patch, the same-valley exchange scattering with momentum transfer $\textbf{Q}^+$ becomes redundent, and the fermionic statistics dictates that the intra-patch density-density interaction vanishes in the infrared limit at the van Hove filling under the patch approximation. We therefore do not consider the intra-patch density-density interaction in Eq. \ref{eq:H6}.  

Next, we study how these inter- and intra-patch interactions $\tilde{g}_p$ 
evolve with a decreasing energy towards the long-wavelength limit. We find that the evolution is governed by the following RG equations up to the one-loop level: 
\begin{align}
&\frac{dg_2}{dy}=-g_2^2-2g_3^2-4d_1(y)g_6g'_6+d_2(y)g_2^2,\nonumber\\
&\frac{dg_3}{dy}=-2g_2g_3-g_3^2+2d_3(y)g_3g'_6+2\tilde{d}_3(y)g_3g_6,\nonumber\\
&\frac{dg_6}{dy}=-d_1(y)(2g_2g'_6+g_6^2+g_6'^2)+\tilde{d}_3(y)(g_3^2+g_6^2)-d_5(y)g_6^2,\nonumber\\
&\frac{dg'_6}{dy}=-2d_1(y)[g_2g_6+g_6g'_6]+d_3(y)(g_3^2+g_6'^2)-\tilde{d}_5(y)g_6'^2,
\label{giRGeqn6}
\end{align}
where $g_p\equiv\nu_0 \tilde{g}_p$ denotes the dimensionless interactions corresponding to the interactions $\tilde{g}_p$ in Eq. \ref{eq:H6}, and $y\equiv \Pi^{n\bar{n}}_{pp}(0)/\nu_0=\frac{1}{2}\rm{ln}^2$$(\frac{\Lambda}{E})$ is the RG running paramater, which is negatively related to the energy $E$. 

Here, we define the following energy-dependent d factors to parametrize the relative magnitudes between different bare susceptibilities and the Cooper instability $\Pi^{n\bar{n}}_{pp}(0)$
\begin{align}
&d_1(y)\equiv\frac{1}{\nu_0}\frac{d\Pi^{nn}_{ph}(0)}{dy},
~~~~~~~~~~~~d_2(y)\equiv\frac{1}{\nu_0}\frac{d\Pi^{n\bar{n}}_{ph}(\textbf{Q}'_{n})}{dy},\nonumber\\
&d_3(y)\equiv\frac{1}{\nu_0}\frac{d\Pi^{n\bar{m}}_{ph}(\textbf{Q}^-_{n\bar{m}})}{dy},
~~~~~~\tilde{d}_3(y)\equiv\frac{1}{\nu_0}\frac{d\Pi^{nm}_{ph}(\textbf{Q}^+_{nm})}{dy},\nonumber\\
&d_4(y)\equiv\frac{1}{\nu_0}\frac{d\Pi^{nn}_{pp}(-\textbf{Q}'_{n})}{dy},
~~~~~~~d_5(y)\equiv\frac{1}{\nu_0}\frac{d\Pi^{nm}_{pp}(\textbf{Q}^-_{\bar{n}m})}{dy},\nonumber\\
&\tilde{d}_5(y)\equiv\frac{1}{\nu_0}\frac{d\Pi^{n\bar{m}}_{pp}(\textbf{Q}^+_{\bar{n}\bar{m}})}{dy}. 
\end{align}
We then model the energy-dependece of these d factors $d_j(y)$, which are generally decreasing functions in $y$, based on their asymptotic behaviors. 
Specifically, in the ultraviolet limit $y\rightarrow 0$, the d factors behave as $d_j(y)\sim 1$, whereas in the infrared limit $y\rightarrow\infty$, they behave as $d_1(y)\sim\frac{1}{\sqrt{2y}}$, $d_2(y)\sim\frac{\gamma_{ph}'}{\sqrt{2y}}$, $d_3(y)\sim\frac{\gamma_{ph}^-}{\sqrt{2y}}$, $\tilde{d}_3(y)\sim\frac{\gamma_{ph}^+}{\sqrt{2y}}$, $d_4(y)\sim\frac{\gamma_{pp}'}{\sqrt{2y}}$, $d_5(y)\sim\frac{\gamma_{pp}^-}{\sqrt{2y}}$, and $\tilde{d}_5(y)\sim\frac{\gamma_{pp}^+}{\sqrt{2y}}$. 
We therefore model the energy dependence of the d factors as $d_1(y)=\frac{1}{\sqrt{1+2y}}$, $d_2(y)=\frac{\gamma_{ph}'}{\sqrt{\gamma_{ph}'^2+2y}}$, $d_3(y)=\frac{\gamma_{ph}^-}{\sqrt{\gamma_{ph}^{-2}+2y}}$, $\tilde{d}_3(y)=\frac{\gamma_{ph}^+}{\sqrt{\gamma_{ph}^{+2}+2y}}$, $d_4(y)=\frac{\gamma_{pp}'}{\sqrt{\gamma_{pp}'^2+2y}}$, $d_5(y) =\frac{\gamma_{pp}^-}{\sqrt{\gamma_{pp}^{-2}+2y}}$, and $\tilde{d}_5(y)=\frac{\gamma_{pp}^+}{\sqrt{\gamma_{pp}^{+2}+2y}}$. 
These d factors characterize important features of low-energy band structures that determine the RG flows. For instance, $d_1(y)$ describes how the density of states evolve with $y$, whereas $d_3(y)$ and $\tilde{d}_3(y)$ describe the evolution of the particle-hole nesting degrees between same-valley patches and opposite-valley patches, respectively.
Similarly, $d_5(y)$ and $\tilde{d}_5(y)$ capture the evolution of the particle-particle nesting degrees 
between the same- and opposite-valley patches, respectively.

By plugging the d factors in the RG equations in Eq. \ref{giRGeqn6} and numerically solving for the evolution of the inter-patch interactions, we find that the relevant interactions $g_p(y)$ flow to the strong coupling limit and diverge as they approaching some critical scale $y_c$. Since $y_c$ corresponds to the critical energy scale at which the perturbative approach breaks down, this energy scale can be associated with the critical temperature $T_c$ at which the FS is destabilized. To quantify the divergent rates of the relevant interactions, we parameterize the interactions in the standard way as  
\begin{align}
g_p(y)=\frac{G_p}{y_c-y}
\label{Gp}
\end{align}
and solve for $G_p$'s. 
In the next section, we will study the dominant instabilities in terms of the effective interaction strengths $G_p$ at $y\rightarrow y_c$. 

\subsection{Two-patch model}
For the two-patch case, the only interaction allowed by the symmetries and spin-valley locking is the inter-valley density-density interaction [see Fig. \ref{gidf}(b)]
\begin{align}
H^{(2)}_{\text{int}}= 
~\tilde{g}\psi^{\dagger}_{K}\psi_{K}\psi^{\dagger}_{K'}\psi_{K'},     
\end{align}
where $\psi^{\dagger}_{\tau}$ creates electrons in the patch at valley $\tau=K,K'$. 
Importantly, since the dispersion is spin-valley locked, the intra-patch density-density interaction vanishes in the infrared limit due to the fermionic statistics, similar to the six-patch case, and the spin-flipping inter-valley exchange scattering is also forbidden. 

The RG equation for this density-density interaction has a simple form up to the one-loop level 
\begin{align}
&\frac{dg}{dy}=-[1-d_{ph}^{\bf Q}(y)]g^2,  
\label{eq:giRGeqn2}
\end{align}
where $g=\bar{\nu}\tilde{g}$ is the dimensionless interaction, and the RG running parameter $y\equiv \frac{\bar{\nu}_{pp}}{\bar{\nu}}|E|^{-1/3}=\Pi_{pp}(\textbf{q}=0,E)/\bar{\nu}$.  
Here, the only d factor that enters the RG equation is $d_{ph}^{\bf Q}(y)\equiv\frac{1}{\bar{\nu}}\frac{d\Pi_{ph}({\bf Q})}{dy}\leq 1$, 
which quantifies how far the particle-hole nesting degree is from perfect. 
In our two-patch model for twisted bilayer TMDs, since the nesting is perfect between the dispersions $\epsilon^K_{\textbf{k}}$ and $\epsilon^{K'}_{\textbf{k}}$ near the van Hove points [see Eq. \ref{eq:dispersion2}], we find that $\Pi_{ph}(\textbf{Q})=\Pi_{pp}(0)$ and thus $d_{ph}^{\bf Q}(y)=1$ [see Appendix \ref{app:bubbles2}].
Consequently, the only allowed interaction $g(y)$ remains marginal and does not diverge in the infrared limit at the one-loop level. 
In systems where the nesting degree deviates from perfect ($d_{ph}^{\bf Q}<1$), the interaction $g(y)$ becomes irrelevant when the bare interaction $g(y=0)>0$ is repulsive. Nonetheless, when the bare interaction $g(y=0)<0$ is attractive, the interaction diverges negatively as $g(y)=\frac{G}{y_c-y}$ when approaching some critical scale $y_c$ with $G$ being the effective attraction.  

\section{Possible instabilities}
With the RG flows of the inter-patch interactions in hand, we can now study the possibile instabilities in the system and identify the most dominant one. 
We do so by first writing down the test vertices $\delta H_i=\Delta_i\psi^{\dagger}\psi^{(\dagger)}+\text{H.c.}$ for instabilities $i$ in both the particle-particle and particle-hole channels [see Fig. \ref{gidf}(c)]. Then by studying the divergence of the RG flows of these vertices in the infrared limit, we can identify the most relevant vertices.  
Importantly, in contrast to the spin-degenerate systems, here the spin-valley locking imposes constraints on what type of instabilities are allowed.  
\begin{figure}[t]
\includegraphics[width=8cm]{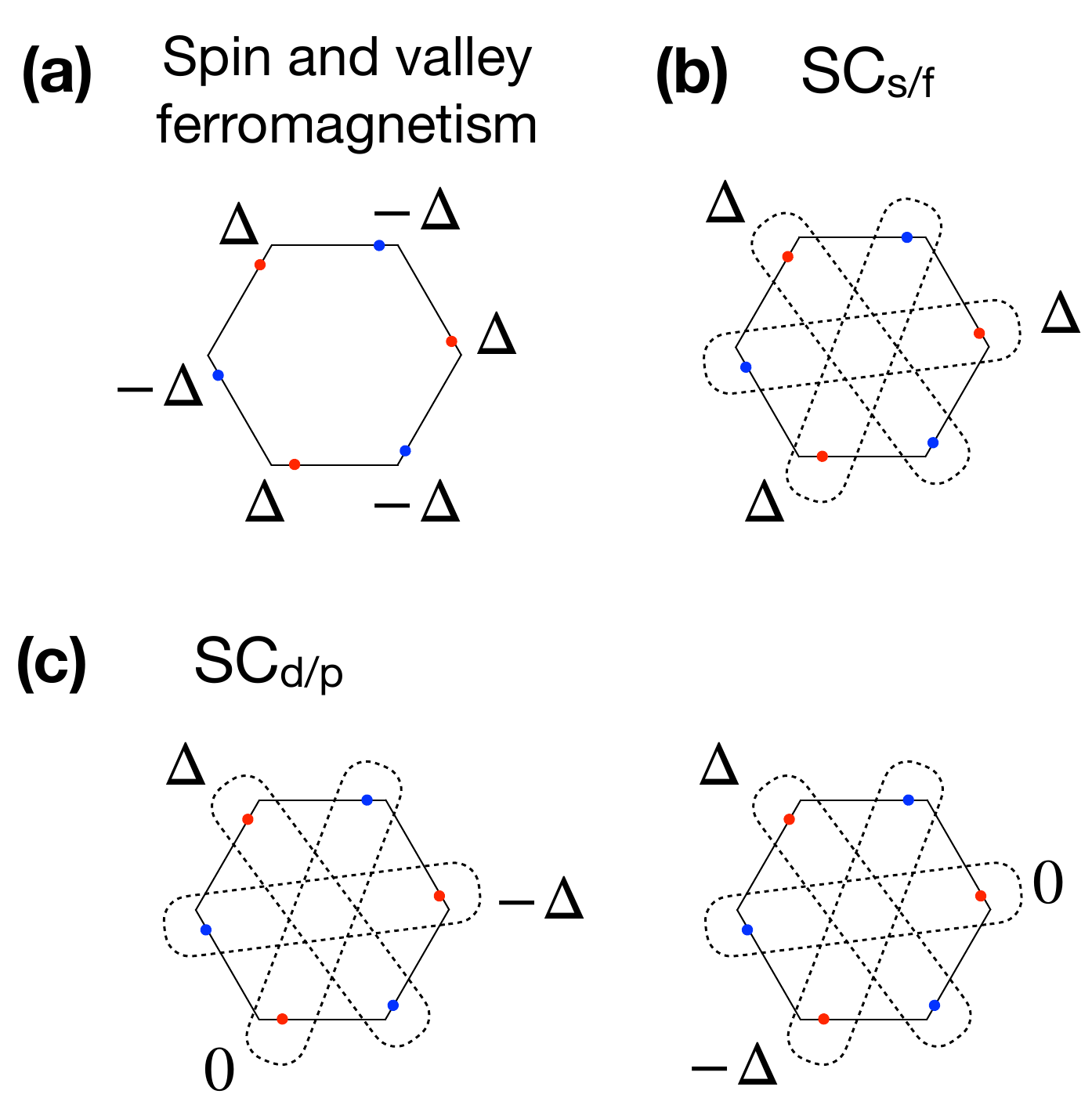}
\caption{Schematic configurations for some of the possible instabilities in the six patch model. (a) Configuration for the spin-valley polarized ferromagnetism, whose tendency is given by $\beta_{FM}^{-}$. (b) Configuration for the $s/f$-wave superconductivity, whose tendency is given by $\beta_{SC}^{s/f}$. (c) The two degenerate configurations for the $d/p$-wave superconductivity, whose tendency is given by $\beta_{SC}^{d/p}$. In all the figures, the hexagon and the red (blue) dots represent the mBZ and the patches from valley $K$ ($K'$), respectively. The dotted lines circle the patches which the corresponding particle-hole or Cooper pairs are from, and $\Delta$ labels the test vertex for each of the instabilities [see Eq. \ref{eq:DeltaFM} and \ref{eq:Deltasc}].}
\label{instability6}
\end{figure}
\subsection{Six-patch model}
We first discuss the six-patch case. For the uniform superconductivity (SC), the test vertex has the form  
\begin{align}
\Delta_{SC}^{n}\psi^{\dagger}_{\bar{n}\bar{s}}\psi^{\dagger}_{ns},
\label{eq:Deltasc}
\end{align}
where $\bar{n}$ labels the opposite patch to patch $n$. The allowed pairing symmetries are then given by different linear combinations of such vertices among different patches [see Fig. \ref{instability6}(b)(c)]. 
Importantly, due to the spin-valley locking, these Cooper pairs have mixed parity since they can be viewed as an equal-mixture of spin-singlet and spin-triplet pairs. For instance, $s$- and $f$-wave are mixed given that they are both fully gapped within the patches, whereas $d$- and $p$-wave are mixed since they share similar nodal structures.  

For pair density waves (PDW) formed by Cooper pairs with finite pair momenta, we consider the following test vertices
\begin{align}
\Delta_{PDW_{a}}^{n}\psi^{\dagger}_{ns}\psi^{\dagger}_{ms},
~~\Delta_{PDW_{b}}^{n}\psi^{\dagger}_{ns}\psi^{\dagger}_{\bar{m}\bar{s}}, 
\end{align}
where $n$ and $m\neq n$ label patches from the same valley.
These two states PDW$_a$ and PDW$_b$ correspond to equal-spin (same-valley) pairs and opposite-spin (opposite-valley) pairs, respectively. Note that the equal-spin PDW that consists of electrons from the same patch is forbidden by fermionic statistics in the infrared limit under the spin-valley locking. 

Next, we discuss the particle-hole instabilities. Due to the spin-valley locking, the magnetic and charge instabilities are generally mixed. Instead, we should consider spin-valley instabilities at zero and finite momentum transfers at $\textbf{Q}'$, $\textbf{Q}^{+}$, and $\textbf{Q}^{-}$. 
Specifically, the test vertex for the spin-valley uniform order is 
\begin{align}
\Delta_{FM}^{n}\psi^{\dagger}_{ns}\psi_{ns},  
\label{eq:DeltaFM} 
\end{align}
and the test vertices for spin-valley density waves with momentum transfer $\textbf{Q}'$, $\textbf{Q}^+$, and $\textbf{Q}^-$ are 
\begin{align}
\Delta_{DW_a}^{n}\psi^{\dagger}_{ns}\psi_{\bar{n}\bar{s}}, 
~~~\Delta_{DW_b}^{n}\psi^{\dagger}_{ns}\psi_{ms}, 
~~~\Delta_{DW_c}^{n}\psi^{\dagger}_{ns}\psi_{\bar{m}\bar{s}},  
\end{align}
respectively, where $n$ and $m\neq n$ label patches from the same valley.
In particular, the density waves with subscripts $a$ and $c$ consist of opposite-valley electron-hole pairs and that with $b$ is an intra-valley density wave. 
Depending on the allowed scattering processes among patches, these instabilities can carry patch-dependent form factors.     

The vertex $\Delta_i$ for each instability $i$ renormalizes with the RG running parameter $y$ as 
$\frac{d\Delta_i}{dy}=-\beta_i\Delta_i$, where $\beta_i=d_{i}\Gamma_i$ quantifies the tendency for instability $i$ to be the most relevant instability. Here, $d_{i}$ and $\Gamma_i$ are the d factor and the driving interaction associated with instability $i$, respectively. In particular, the interaction $\Gamma_i$ for a certain instability $i$ is given by a certain linear combination of the four inter-patch interactions $\{g_p\}$, and can be expressed in terms of the interaction strengths $\{G_p\}$ defined in Eq. \ref{Gp}. 
Furthermore, the renormalization of the susceptibility $\chi_i$ for a given instability $i$ is given by $\frac{d\chi_i}{dy}=d_i|\Delta_i|^2$, which explicitly depends on the evolution of the test vertex $\Delta_i(y)$ [see Fig. \ref{gidf}(d)]. The asymptotic behavior of the susceptibility is therefore controlled by the tendency $\beta_i$ through $\chi_i(y)\sim(y_c-y)^{\alpha_i}$, where $\alpha_i=2\beta_i+1$\cite{RG_FeSc_Chubukov_PRB,RG_Moire_Nandkishore}. 
Since the susceptibility only diverges near the critical scale $y\rightarrow y_c$ when $\alpha<0$, only instabilities with $\beta_i<-1/2$ are relevant and  
the magnitude $|\beta_i|$ controls the diverging rate of the susceptibility. 
Specifically, among the relevant instabilities with $\beta_i<-1/2$, the instability $i$ with the most negative $\beta_i$ dominates in the long-wavelength limit.   

In the following, we use the tendency $\beta_i$ at $y\rightarrow y_c$ as the measure to analyze the competition among instabilities. For the six-patch case, we find that the tendencies for particle-particle instabilities are 
\begin{align}
&\beta_{SC}^{s/f}=G_2+2G_3,~\beta^{d/p}_{SC}=G_2-G_3\nonumber\\
&\beta_{PDW_a}=d_5(y_c)G_6,~
\beta_{PDW_b}=\tilde{d}_5(y_c)G'_6,
\label{betapp6}
\end{align}
where the superscripts label different pairing symmetries. 
In particular, the uniform superconductivity $\beta_{SC}^{s/f}$ and $\beta_{SC}^{d/p}$ have mixed parity [see Fig. \ref{instability6}(b)(c)]. Importantly, $\beta_{SC}^{d/p}$ is two-fold degenerate and the corresponding $d/p$-wave patch configurations are shown in Fig. \ref{instability6}(c). 

For the particle-hole instabilities, we find that 
\begin{align}
&\beta^s_{FM}=-d_1(y_c)(-2G_6-G_2-2G'_6),\nonumber\\
&\beta^f_{FM}=-d_1(y_c)(-2G_6+G_2+2G'_6),\nonumber\\
&\beta^d_{FM}=-d_1(y_c)(G_6-G_2+G'_6),\nonumber\\
&\beta^p_{FM}=-d_1(y_c)(G_6+G_2-G'_6),\nonumber\\
&\beta_{DW_a}=-d_2(y_c)G_2,\nonumber\\
&\beta_{DW_b}^{\pm}=-\tilde{d}_3(y_c)(G_6\mp G_3),\nonumber\\
&\beta_{DW_c}^{\pm}=-d_3(y_c)(G'_6\pm G_3).  
\label{betaph6}
\end{align}
The supercript in each tendency labels the patch-dependent symmetry form factors of the instability. For instance, the $f$-wave spin-valley uniform order corresponds to the spin- and valley-polarized ferromagnetism [see Fig. \ref{instability6}(a)]. 

\subsection{Two-patch model}
For the two-patch case, the instabilities allowed by the spin-valley locking are the uniform superconductivity with mixed-parity, the spin-valley polarized ferromagnetism, and the spin-valley density wave. The corresponding test vertices are $\Delta_{SC}\psi^{\dagger}_{n}\psi^{\dagger}_{\bar{n}}$, $\Delta_{FM}\psi^{\dagger}_{n}\psi_{n}$, and $\Delta_{DW}\psi^{\dagger}_{n}\psi_{\bar{n}}$, respectively, where $n=K,K'$ labels the two patches [see Fig. \ref{FS}(f)] and $\bar{n}$ labels the other patch. 
The tendencies of these instabilities are given by  
\begin{align}
&\beta_{SC}=G,
~~~~\beta_{FM}^{\pm}=0,\nonumber\\
&\beta_{DW}=-d_{ph}^{\bf Q}(y_c)G,\nonumber\\
\label{beta2}
\end{align}
respectively, where $d_{ph}^{\bf Q}(y_c)=1$ for the dispersions in Eq. \ref{eq:dispersion2} due to the perfectly nested Fermi surface, and $d_{ph}^{\bf Q}(y_c)<1$ when the Fermi surface deviates from the perfect nesting limit. Importantly, the spin-valley uniform orders, which include an overall chemical potential shift ($FM^{+}$) and spin-valley polarized ferromagnetism ($FM^{-}$), have zero tendency in becoming the dominant instability. This is because these instabilities are driven by the intra-patch density-density interaction, which is forbidden by the fermionic statistics in the infrared limit under the spin-valley locking. In contrast, the superconductivity and spin-valley density wave are expected to dominate in the presence of an attractive and repulsive inter-valley density-density interaction $G$, respectively.  

\begin{figure}[t]
\includegraphics[width=8cm]{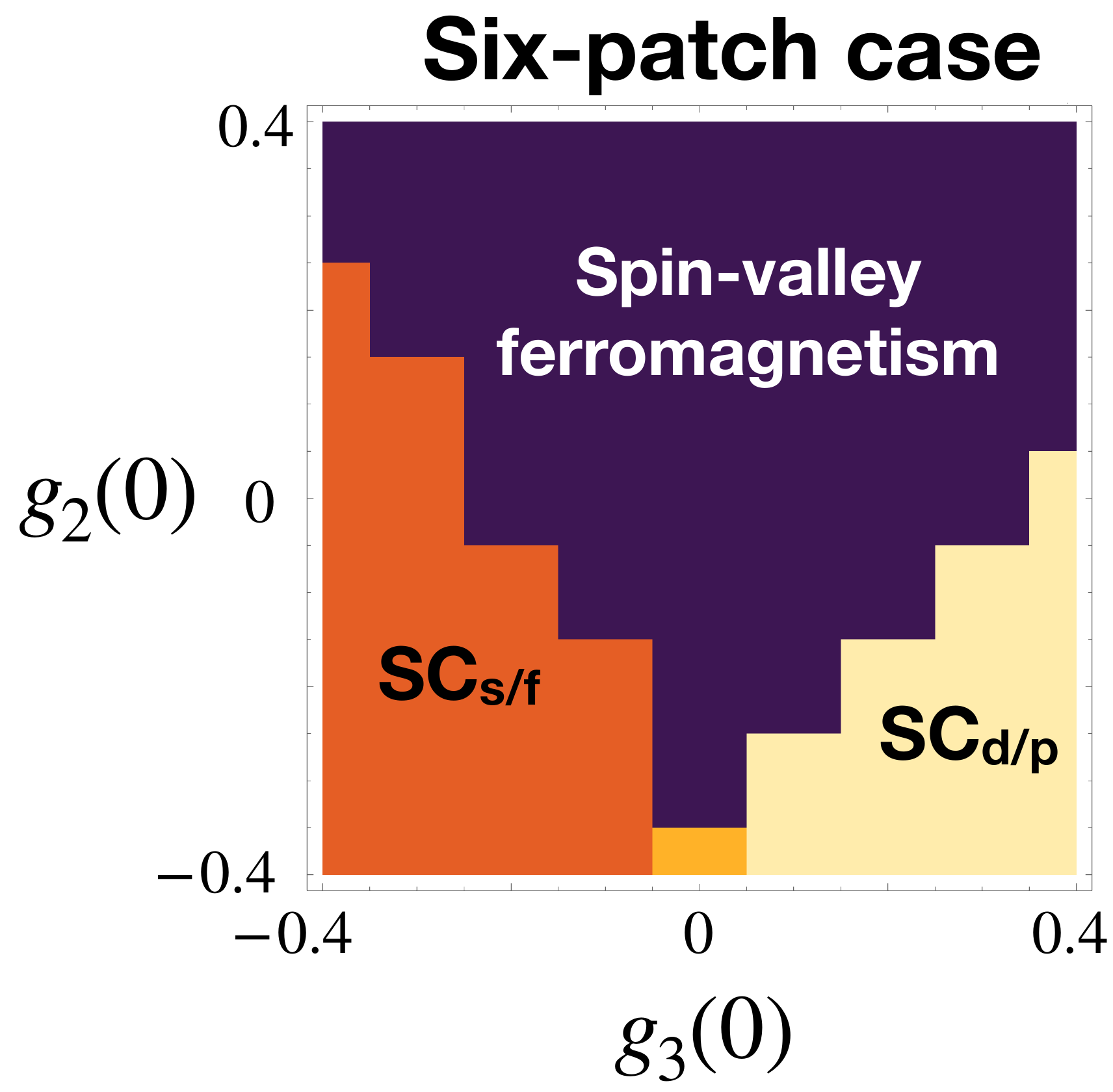}
\caption{The phase diagram for the six-patch model. In the dark yellow regime between the $s/f$-wave superconductivity (SC$_{s/f}$) and the $d/p$-wave superconductivity (SC$_{d/p}$), the two superconducting phases are degenerate and dominant.  
Here we set the bare interactions $g_6(y=0)=-0.2$, $g_6'(y=0)=0.2$, and the nesting parameters $\gamma_{ph}^-=\gamma_{pp}^-=\gamma_{pp}'=1$, $\gamma_{ph}^+=\gamma_{pp}^+=\gamma_{ph}'=0.8$.}
\label{PD6}
\end{figure}
\section{phase diagram}\label{sec:PD}
Using the perturbative RG method described above, in this section we investigate the most relevant  instabilities when tuning the signs and magnitudes of the bare interactions $g_p(y=0)$ for both the six- and two-patch models.

\subsection{Six-patch model}
We study the dominant instabilities under attractive and repulsive bare inter-patch interactions $g_2(y=0)$, $g_6'(y=0)$, $g_6(y=0)$, and $g_3(y=0)$. Specifically, the former two and the third are inter- and intra-valley density-density interactions, respectively, and the last one is an intra-valley scattering of a zero-momentum pair. Given the dispersions in the six-patch case [see Eq. \ref{eq:dispersion2} and Fig. \ref{FS}(a)], we will focus on the weak nesting regime where we set $\gamma_{ph}^-=\gamma_{pp}^-=\gamma_{pp}'=1$ and $\gamma_{ph}^+=\gamma_{pp}^+=\gamma_{ph}'=0.8$. The nesting degrees between inter- and intra-valley patches are in general different due to the curvatures of the Fermi surface near the patches. 

We find that uniform superconductivity and spin-valley ferromagnetism dominate the phase space up to a finite interaction strength away from the infinitesimal limit, where we focus on the range $|g_p(y=0)|\leq 0.2$.  
This is in contrast to spin-degenerate six-patch systems, such as twisted double bilayer graphene, where the intra-pocket pair density wave (PDW) and charge density wave (CDW) states also dominate over a substantial portion of the phase space when the inter-valley scattering is negligible\cite{DBLG_RG_Hsu}. 
In fact, the suppression of PDW and CDW in twisted bilayer TMDs we found here is a direct consequence of the spin-valley locking.  
Specifically, due to the spin-valley locked Fermi surface, the intra-pocket PDW is forbidden by the fermionic statistics and the CDW mixes with a less favorable spin density wave into a spin-valley density wave that does not dominate in the considered ranges of interaction strength and nesting degree. 

Here we show the phase diagram in Fig. \ref{PD6} to demonstrate how the dominant instabilities change under repulsive and attractive $g_2(y=0)$ and $g_3(y=0)$ in the presence of attractive $g_6(y=0)$ and repulsive $g_6'(y=0)$. We choose to show this phase diagram because it serves as a representative that contains both the dominant superconducting and particle-hole instabilities within the considered parameter range. 

\begin{figure}[t]
\includegraphics[width=8cm]{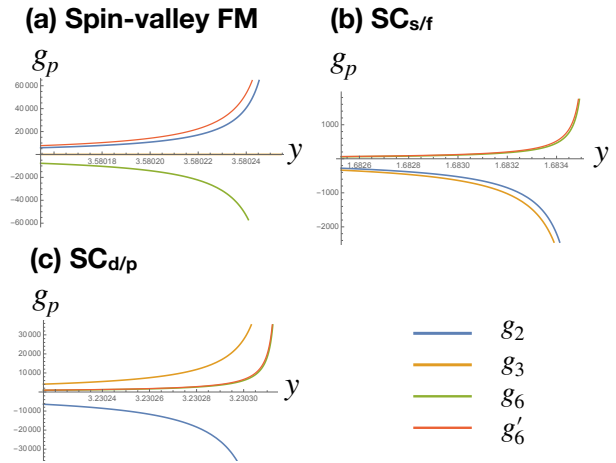}
\caption{Representative RG flows of inter-patch interactions $g_p$ in the regimes where
the dominant instabilities are the (a) spin and valley polarized ferromagnetism, (b) $s/f$-wave superconductivity, and (c) $d/p$-wave superconductivity.}
\label{giflow6}
\end{figure}

We first discuss the lower half of the phase diagram.   
Generally speaking, we find that an attractive density-density interaction $g_2$ between opposite momenta patches promotes uniform superconductivity while the inter-patch pair scattering $g_3$ drives a superconducting phase transition from the $s/f$-wave pairing to the $d/p$-wave pairing. 
The inter-patch pair scattering $g_3$ is the key parameter that drives the transition because the momentum transfer in this scattering process corresponds to the sign change in the $d/p$-wave superconducting gap [see Fig. \ref{instability6}(c)]. 
Specifically, an attractive and repulsive $g_3$ promotes the $s/f$-wave pairing without sign change and the $d/p$-wave pairing with sign change among patches, respectively [see Eq. \ref{betapp6}]. 
Furthermore, it is clear from the dominante term $-2g_2g_3$ in the RG equation of $g_3$ that the bare scattering $g_3(y=0)>0$ ($<0$) leads to a repulsively (attractively) relevant $g_3$ when the density-density interaction $g_2$ is attractive.
This gives rise to the fixed trajectories we find for the two superconducting phases in Fig. \ref{giflow6}(b) and (c), and thus the lower half of the phase diagram in Fig. \ref{PD6}.  
Importantly, this $d/p$-wave pairing we find has two degenerate configurations [see Fig. \ref{instability6}(c)] such that we expect it to be a topological chiral paired state due to energetic reasons. 

We now turn to the upper half of the phase diagram.  
We find that the spin-valley polarized ferromagnetism is promoted by a repulsive inter-valley density-density interaction $g_2$ and dominates over density waves even up to an intermediate nesting degree $\gamma^-=\gamma^+=10$. 
This is mainly because the driving interactions $\Gamma_{DW_i}$ for the density waves is suppressed more by the spin-valley locking than those for the uniform particle-hole phases $\Gamma_{FM}^i$. 
The tendency for the density waves $\beta_{DW_i}=d_i\Gamma_{DW_i}$ therefore remains subdominant even with an intermediate inter-valley (intra-valley) nesting d factor $d_3$ ($\tilde{d}_3$).  
Specifically, the density waves rely heavily on the inter-patch scattering $g_3$, whose momentum transfer is related to the modulating $\textbf{q}$ of the density waves, whereas the uniform phases depend only on the density-density interactions [see Eq. \ref{betaph6}]. 
Since the density-density interactions are generally more relevant than the scattering term $g_3$, especially the one between opposite patches ($g_2$), within the parameter range we study the uniform phases always dominate over the density waves. 

Within all the uniform phases, we find that the spin- and valley-polarized ferromagnetism, whose tendency is given by $\beta_{FM}^f$ in Eq. \ref{betaph6}, is the most dominant [see Fig. \ref{PD6}]. 
The sign of the order parameter in this spin and valley polarized phase alternates between the two oppositely spin-polarized valleys [see Fig. \ref{instability6}(a)]. This spin-valley ferromagnetism therefore dominates when the inter-valley density-density interactions $g_2$ and $g_6'$ are repulsive and the intra-valley density-density interaction $g_6$ is attractive [see Fig. \ref{giflow6}(a)]]. 

\begin{figure}[t]
\includegraphics[width=8cm]{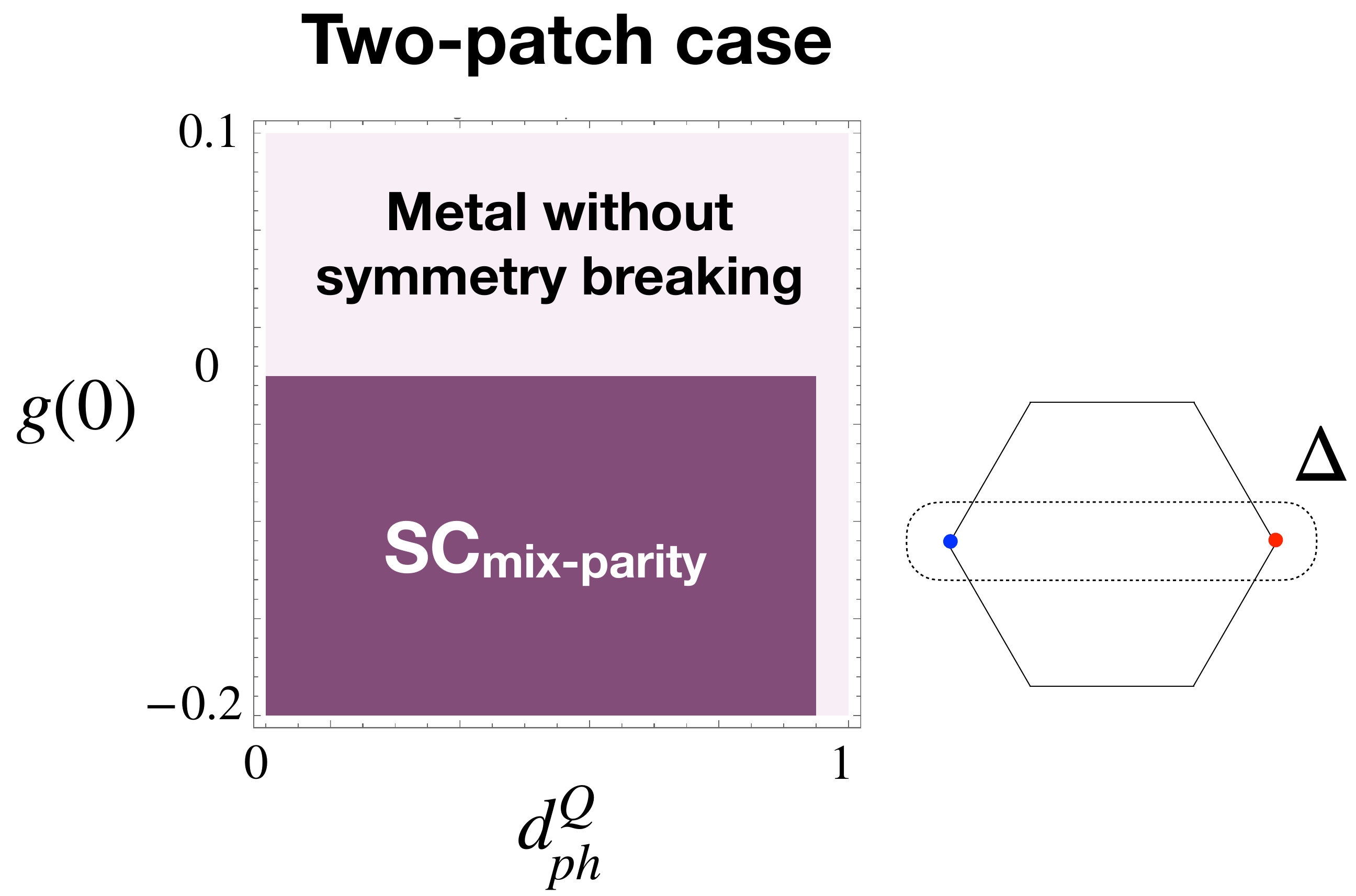}
\caption{The left panel shows the phase diagram for the two-patch model. The d factor $d_{ph}^{\textbf{Q}}\in[0,1]$ parameterizes the nesting degree between the two patches, and $d_{ph}^{\textbf{Q}}=1$ in the perfect nesting limit. The left panel shows the schematic configuration for the mixed parity superconductivity. The red (blue) dot represents the patch for the higher-order VHS from valley $K$ ($K'$), respectively. The hexagon represents the mBZ, the dotted line circles the patches which the Cooper pairs are from, and $\Delta$ labels the superconducting test vertex [see the paragraph above Eq. \ref{beta2}].}
\label{PD2}
\end{figure}
\subsection{Two-patch model}
For the two-patch case, the results strongly depend on whether the particle-hole nesting degree is perfect or not due to the spin-valley locking. 
While the dispersion $\epsilon_{\textbf{k}}^{K/K'}$ we use in Eq. \ref{eq:dispersion2} exhibits perfect nesting, more realistic dispersions for certain materials may have curvatures on the Fermi surface and exhibit deviation from perfect nesting. 
Such curvature can occur when higher-order terms in momentum $O(k^4)$ are included in the dispersions. We will therefore discuss the expected phases in both the perfect nesting limit and the non-perfect nesting cases.

The phase diagram is shown in Fig. \ref{PD2}. 
In the perfect nesting limit where $d_{ph}^{\textbf{Q}}=1$, although the bare susceptibilities for superconductivity as well as the particle-hole instabilities are all diverging, we find that none of the driving interactions diverge in the low-energy limit up to one-loop order. 
This is clear from the RG equation of the only interaction $g$ that survives the spin-valley locking [see Eq. \ref{eq:giRGeqn2}]. When the nesting is perfect, the d factor $d_{ph}^{\textbf{Q}}=1$ such that $g$ is marginal at one-loop level and does not diverge in the long-wavelength limit. 
Since all the allowed instabilities are driven by either an attractive or repulsive $g$ [see Eq. \ref{beta2}]
, we find a marginal metallic state without symmetry breaking in the perfect nesting limit. 
Note that this metallic state with diverging susceptibilities but without symmetry breakings only exists because (1) the spin-valley locking suppresses the existence of other inter- and intra-patch interactions besides $g$ such the RG equation takes the simple form in Eq. \ref{eq:giRGeqn2}, and (2) the nesting degree is perfect. 
To determine whether this spin-valley-locking-induced metallic state survives higher-order perturbations, higher-loop calculations are necessary. 

When the Fermi surface deviates from the perfect nesting limit, the corresponding d factor $d_{ph}^{\textbf{Q}}<1$ deviates from unity such that the inter-patch density-density interaction $g$ is not marginal anymore. 
Instead, $g$ becomes irrelevant when the bare interaction $g(y=0)$ is repulsive, and becomes a relevant attraction when the bare interaction $g(y=0)$ is attractive.
This suggests that when the bare interaction is repulsive, the driving repulsion for the spin-valley density wave is irrelevant such that we again expect a metallic state without symmetry breaking.   
In contrast, when the bare interaction is attractive, $g$ becomes a relevant attraction such that superconductivity is dominant. Superconductivity therefore becomes the only possible instability when there are two higher-order VHS from opposite valleys lying at the chemical potential and when the Fermi surface near the VH points deviates from the perfect nesting limit. 

In both the perfect and non-perfect nesting cases, we dub the two spin-valley-locking induced metallic states as  `supermetal states' [see Fig. \ref{PD2}] because similar to the supermetals previously found in systems with a single\cite{supermetal} and six \cite{RG_HOVH_Scherer} spin-degenerate higher-order VHS, they have diverging bare susceptibilities but develop no symmetry breakings in the infrared limit. 
Nonetheless, in contrast to the previously found supermetal states, the metallic states we find occur because the driving interactions are marginal or irrelevant and are still Fermi-liquid metals instead of a non-Fermi liquid state. 

\section{Summary and discussion}\label{sec:summary}
In this work, we apply a perturbative renormalization group method to study the dominant Fermi surface instabilities in twisted homobilayer transition metal dichalcogenides (TMD). 
Such type of moir\'e systems feature a spin-valley locked low-energy band structure due to the spin-orbit coupling as well as two types of van Hove singularities (VHS) near the Fermi surface, the conventional ones and the higher-order ones. The density of states exhibit logarithmic divergence and power-law divergence with an exponent of $1/3$ in the former and latter cases, respectively. 
By tuning the strength of an applied displacement field, there are either six conventional or two higher-order VHS. We therefore consider `hot-spot' type models for each of the situations and study models with six and two patches around the van Hove singularities. 

For the six-patch model, we find that mixed parity superconductivity and spin-valley polarized ferromagnetism are dominant, depending on the signs of density-density interactions. In contrast to the results from spin-degenerate cases\cite{DBLG_RG_Hsu}, where density waves dominate a substantial portion of the phase space, here the spin-valley locking suppresses the density waves through symmetry constraints on the allowed interactions and instabilities. 
Moreover, we find that a superconducting phase transition can be driven by an intra-valley pair scattering from an $s/f$-wave paired state to a chiral $d/p$-wave topological paired state. 

For the two-patch model, we find that whether the nesting degree is perfect or not plays an important role in the symmetry-breaking pattern under the effects of spin-valley locking. 
In particular, we find a metallic state without symmetry breaking in the perfect nesting limit due to the marginal driving interaction at one-loop level. Further studies are needed to determine whether this marginal metal state survives under higher-loop perturbations in the RG analysis. 
In contrast, for non-perfectly nested dispersions, we find superconductivity and another metallic state without symmetry breaking due to irrelevant interactions, depending on whether the density-density interaction is attractive or repulsive. 
This mixed parity superconductivity is the only possible instability when there are two spin-valley locked higher-order van Hove singularities on the Fermi surface, and the suppression of the particle-hole instabilities is due to the suppression of other inter- and intra-valley interactions by the spin-valley locking.    

We make four comments about our results. 
First, the superconducting phases we obtained in this work using the parquet RG method can be viewed as the results of a generalized lattice version of Kohn-Luttinger mechanism. 
In particular, although we do not show the phase diagrams here, we have checked for the six-patch model the case where all four bare interactions $g_p(y=0)$ are repulsive. In such a case we find $d/p$-wave superconductivity. 
The main difference between the parquet RG method and considering just the Kohn-Luttinger superconductivity is that the symmetry-allowed particle-hole instabilities are also considered in the competition and that the fluctuations arising from these particle-hole instabilities can also contribute to the effective attractions that drive the superconductivity.

Second, we comment on the relations between the bare interactions $g_p(y=0)$ and the microscopic interactions. In this RG calculation, the ultraviolet limit $\Lambda$ that we consider is in fact not given by the energy scale of the band width. Instead, $\Lambda$ is given by the energy scale associated with the patch size, which is much smaller than the band width. Consequently, the bare inter-patch interactions $g_p(y=0)$ do not directly correspond to the microscopic interactions.
In order to pinpoint the type of microscopic interactions that lead to a certain set of repulsive or attractive inter-patch interactions $g_i(y=0)$, one will need to study the RG flows from the band width to $\Lambda$ for the microscopic interactions of interest, such as the screened Coulomb interaction, electron-phonon couplings, or spin-valley fluctuations. This is out of the scope of this work and is left as an interesting future direction. 

Third, for cases with higher-order VHS where the density of states is power-law divergent, the parquet RG method is strictly speaking an approximation for the direct perturbative diagrammatic technique when going beyond the one-loop order. 
Nonetheless, Ref. \onlinecite{RG_HOVH_Scherer} has found that such an approximation is a good one for a specific case of power-law diverging density of states with an exponent $\eta=1/4$. 
Specifically, Ref. \onlinecite{RG_HOVH_Scherer} has shown explicitly for the two-loop order and argued for higher-loop orders that compared to the direct diagrammatic result, the RG result for the intra-patch interaction qualitatively captures the temperature dependence and is only quantitatively off by an $O(1)$ prefactor.
In contrast, when the density of states is logarithmically diverging (conventional VHS), the RG method and the diagrammatic technique results match perfectly at any order.
For our two-patch case with higher-order VHS, although the density of states diverges with a different exponent $\eta=1/3$, we do not expect qualitative difference from the $\eta=1/4$ case. 
Moreover, since we only consider corrections up to one-loop order, we expect the RG results to match with the direct diagrammatic results for both the two-patch and six-patch models. 
Further careful investigations into the difference between the RG and the direct diagrammatic results beyond the one-loop order for our $\eta=1/3$ two-patch case will be interesting future works. 

Finally, since the effective interaction strength in twisted homobilayer TMD can be tuned by the twist angle and the displacement field experimentally, we expect that the weak-coupling physics we study in this work can be accessed in experiments. 
Moreover, although direct manipulation of the signs of the inter-valley interactions might not be accessible, given the rich variety of stable compounds and lattice structures of transition metal dichalcogenides, 
we expect that the phases we find can be observed experimentally in various twist homobilayer TMD compounds. 

\emph{Acknowledgment}--- The authors thank Andrey Chubukov for helpful discussions. We acknowledge the support by the Laboratory for Physical Sciences.
%

\appendix
\section{The density of states and bare susceptibilities in the two-patch case}
\label{app:bubbles2}
For the case with two higher-order VHS, the low-energy dispersion near the VH points from the two valleys are given in Eq. \ref{eq:dispersion1} in the main text 
\begin{align}
&\epsilon_{\textbf{k}}^K=\kappa(k_x^3-3k_xk_y^2)\nonumber\\
&\epsilon_{\textbf{k}}^{K'}=-\kappa(k_x^3-3k_xk_y^2),
\label{eq:A1}
\end{align}
where $\kappa$ is given by the overall energy scale. 
\subsection{Density of states}
The density of states per patch is therefore given by
\begin{align}
\nu(E)&=\frac{1}{(2\pi)^2}\int dk_xdk_y\delta(E-\epsilon^K_k)\nonumber\\
&=\frac{1}{(2\pi)^2\sqrt{3}|\kappa|}\int dk_x\frac{1}{\sqrt{k_x(k_x^3-E/\kappa)}}. 
\end{align}
For $E/\kappa>0$,
\begin{align}
&\int dk_x\frac{1}{\sqrt{k_x(k_x^3-E/\kappa)}}\nonumber\\
&=(\int_{-\infty}^0+\int_{E^{\frac{1}{3}}}^\infty)dk_x\frac{1}{\sqrt{k_x(k_x^3-E/\kappa)}}
=\frac{\sqrt{\pi}|\kappa|^{\frac{1}{3}}}{|E|^{\frac{1}{3}}}\frac{\Gamma(\frac{1}{3})}{\Gamma(\frac{5}{6})}. 
\end{align}
For $E/\kappa<0$, we obtain the same result. The density of states is thus given by 
\begin{align}
\nu(E)&=\frac{1}{4\sqrt{3}\pi^{\frac{3}{2}}}\frac{\Gamma(\frac{1}{3})}{\Gamma(\frac{5}{6})}|\kappa|^{-\frac{2}{3}}|E|^{-\frac{1}{3}}\equiv\bar{\nu}|E|^{-\frac{1}{3}}. 
\end{align}

\subsection{$\Pi_{ph}(0)$}
The bare particle-hole susceptibility at $\textbf{q}=0$ is given by 
\begin{align}
\Pi_{ph}(0)&=-\text{lim}_{\textbf{q}\rightarrow 0}\int d\textbf{k}\frac{f_{\epsilon^n_{\textbf{k}}}
-f_{\epsilon^n_{\textbf{k}+\textbf{q}}}}{\epsilon^n_{\textbf{k}}-\epsilon^n_{\textbf{k}+\textbf{q}}}\nonumber\\
&=-\int d\epsilon\nu(\epsilon)\frac{\partial f}{\partial\epsilon}
=\frac{\bar{\nu}}{T}\int_{-\Lambda}^{\Lambda} d\epsilon|\epsilon|^{-1/3}\cosh^{-2}(\beta\epsilon/2)\nonumber\\
&=\frac{\bar{\nu}}{T^{1/3}}\frac{1}{4}\int_{-\Lambda}^{\Lambda} d\epsilon|\epsilon|^{-1/3}\cosh^{-2}(\epsilon/2)\nonumber\\
&=\frac{\bar{\nu}_{ph}}{T^{1/3}}, 
\end{align}
where $n=K,K'$ labels the valley, $T$ is the temperature, $\beta=1/T$, the ultraviolet scale $\Lambda$ is given by the scale of the patch size, and $\bar{\nu}_{ph}\equiv\alpha\bar{\nu}$ with $\alpha=\frac{1}{4}\int_{-\Lambda}^{\Lambda} d\epsilon|\epsilon|^{-1/3}\cosh^{-2}(\epsilon/2)\sim 1.14$. 

\subsection{$\Pi_{pp}(0)$}
The Cooper is given by 
\begin{align}
\Pi_{pp}(0)&=\text{lim}_{\textbf{q}\rightarrow 0}\int d\textbf{k}\frac{1-f_{\epsilon^K_{\textbf{k}}}
-f_{\epsilon^{K'}_{-\textbf{k}+\textbf{q}}}}{\epsilon^K_{\textbf{k}}+\epsilon^{K'}_{-\textbf{k}+\textbf{q}}}\nonumber\\
&=\int d\epsilon\nu(\epsilon)\frac{1-2f_{\epsilon^K}}{2\epsilon^K}
=\bar{\nu}\int_{-\Lambda}^{\Lambda} d\epsilon|\epsilon|^{-1/3}\frac{1}{2\epsilon}\tanh(\beta\epsilon/2)\nonumber\\
&=\frac{\bar{\nu}}{T^{1/3}}\frac{1}{2}\int_{-\Lambda}^{\Lambda} d\epsilon|\epsilon|^{-4/3}\tanh(\epsilon/2)\nonumber\\
&=\frac{\bar{\nu}_{pp}}{T^{1/3}}, 
\end{align}
where $\bar{\nu}_{pp}\equiv\tilde{\alpha}\bar{\nu}$ with $\tilde{\alpha}=\frac{1}{2}\int_{-\Lambda}^{\Lambda} d\epsilon|\epsilon|^{-4/3}\tanh(\epsilon/2)\sim 3.4$. 

\subsection{$\Pi_{ph}(\textbf{Q})$}
From Eq. \ref{eq:A1}, it is clear that the dispersions near the two VHS satisfy $\epsilon^K_{\textbf{k}}=-\epsilon^{K'}_{\textbf{k}}$, where the $\textbf{k}$ in $\epsilon^K_{\textbf{k}}$ and $\epsilon^{K'}_{\textbf{k}}$ are the relative momenta measured from the two van Hove points, respectively. 
The Fermi surface within the two patches therefore satisfies the inter-valley perfect nesting condition $\epsilon^{K}_{\textbf{p}}=-\epsilon^{K'}_{\textbf{p}+\textbf{Q}}$, where $\textbf{p}$ is the momentum measured from $\Gamma$, and $\textbf{Q}$ is the momentum that connects the two van Hove points [see Fig. \ref{FS}(d) in the main text]. The particle-hole bare susceptibility at $\textbf{q}=\textbf{Q}$ is therefore given by 
\begin{align}
\Pi_{ph}(\textbf{Q})&=-\int d\textbf{p}\frac{f_{\epsilon^K_{\textbf{p}}}
-f_{\epsilon^{K'}_{\textbf{p}+\textbf{Q}}}}{\epsilon^{K}_{\textbf{p}}-\epsilon^{K'}_{\textbf{p}+\textbf{Q}}}
=-\int d\textbf{p}\frac{f_{\epsilon^K_{\textbf{p}}}
-f_{-\epsilon^{K}_{\textbf{p}}}}{2\epsilon^{K}_{\textbf{p}}}\nonumber\\
&=\int d\epsilon\nu(\epsilon)\frac{1-2f_{\epsilon^K}}{2\epsilon^K}\nonumber\\
&=\Pi_{pp}(0). 
\end{align}
The d factor $d_{ph}^{\bf Q}(y)\equiv\frac{1}{\bar{\nu}}\frac{d\Pi_{ph}({\bf Q})}{dy}$ that quantifies the particle-hole nesting degree is therefore 1 since we choose the RG running parameter to be $y=\Pi_{pp}(0)/\bar{\nu}$.

\end{document}